\documentclass[conference]{IEEEtran}
\usepackage{amsmath,amssymb,amsfonts}
\usepackage{amsmath,amssymb,dsfont,bbm, float, tabularx, algorithm2e}

\usepackage{cite}
\usepackage{hyperref}

\usepackage{algorithmic}
\usepackage{graphicx}
\usepackage{textcomp}
\usepackage{xcolor}
\usepackage{subcaption}
\def\BibTeX{{\rm B\kern-.05em{\sc i\kern-.025em b}\kern-.08em
    T\kern-.1667em\lower.7ex\hbox{E}\kern-.125emX}}

\usepackage{url}

\begin{document}
	
	\title{Scalable and Privacy-Preserving Synthetic Data Generation on Decentralised Web}
	
	\author{\IEEEauthorblockN{Vishal Ramesh}
		\IEEEauthorblockA{\textit{University of Oxford} \\
			Oxford, United Kingdom\\
			vishal.ramesh@hertford.ox.ac.uk}
		\and
		\IEEEauthorblockN{Rui Zhao \textsuperscript{\textsection}}
		\IEEEauthorblockA{\textit{University of Oxford} \\
			Oxford, United Kingdom\\
			rui.zhao@cs.ox.ac.uk}
		\and
		\IEEEauthorblockN{Naman Goel \textsuperscript{\textsection}}
		\IEEEauthorblockA{\textit{University of Oxford and Alan Turing Institute} \\
			Oxford, United Kingdom\\
			naman.goel@alumni.epfl.ch}
	}
	
	\maketitle
	\begingroup\renewcommand\thefootnote{\textsection}
	\footnotetext{Corresponding Authors}
	\endgroup
	
	\begin{abstract}
		Data on the Web has fueled much of the recent progress in AI. As more high-quality data becomes difficult to access, synthetic data is emerging as a promising solution for privacy-friendly data release and complementing real datasets in developing robust and safe AI. But there is limited work on decentralised, scalable and contributor-centric synthetic data generation systems. A recent proposal, called Libertas~\cite{oxford}, allows data contributors to autonomously participate in joint computations over their Web data without relying on a trusted centre. Libertas uses Solid (Social Linked Data) and MPC (Secure Multi-Party Computation) to achieve this goal. Solid is a decentralised Web specification that lets anyone store their data securely in their personal decentralised data stores called Pods and control which applications have access to their data. MPC refers to the set of cryptographic methods for different parties to jointly compute a function over their inputs while keeping those inputs private. Thus, Libertas can also be used to generate synthetic data from otherwise inaccessible Web data in a responsible way, by ensuring contributor autonomy, decentralisation and privacy. However, the scalability of this system remains limited due to the high computation and communication costs in MPC. In this paper, we show how one can improve Libertas using secure enclaves (in addition to MPC) to address the scalability challenge. Secure enclaves such as Intel SGX rely on hardware based features for confidentiality and integrity of code and data. We discuss a principled approach for integrating SGX within the Libertas architecture for scalable differentially private synthetic data generation, and support our analysis with rigorous empirical results on simulated and real datasets and different synthetic data generation algorithms.
	\end{abstract}
	

\section{Introduction}
In today’s data-driven world, the need for large and diverse datasets is greater than ever. These datasets are instrumental in training machine learning models, conducting research, and fuelling innovation across various domains, from healthcare to finance and beyond. However, the acquisition and sharing of real-world data in a responsible way have become increasingly challenging due to concerns about privacy, security and the potential misuse of sensitive information. 

In response to these challenges, synthetic data generation has emerged as a promising solution~\cite{jordon2022synthetic}. Synthetic data refers to artificially generated data that mimics the statistical properties of real-world data. It offers a way to balance the need for data-driven insights and open availability with protecting individual privacy~\cite{emam_practical_2020}. Beyond privacy, synthetic data approaches are also being actively explored to overcome the limitations and shortcomings of real data for building more robust artificial intelligence~\cite{biasforbes}.

There exists a rich body of research dedicated to the algorithmic intricacies of synthetic data generation~\cite{hu2023sok}. The focus remains on development of novel algorithms for generating synthetic data with properties closely mirroring those of real-world data, addressing challenges such as preserving statistical distributions, correlations, and structural features while guaranteeing individual privacy. However, while these algorithmic aspects are crucial, for synthetic data to be truly responsible, trustworthy, and privacy-preserving, it is necessary to take a holistic perspective. This encompasses not only the algorithms used for data synthesis but also the entire lifecycle of data, starting with the contributors of the real data. Contributor autonomy should be one of the central tenets of responsible synthetic data generation. Contributors, who provide the real data, should have a significant degree of control and decision-making power throughout the synthetic data generation process. This autonomy ensures that their interests and concerns are taken into account, ultimately promoting trust and ethical data handling practices.

Further, a common assumption in most synthetic data generation approaches is the existence of a central curator. There are several problems with this approach. It expects diverse set of contributors to trust a common party to manage their (real) data, which may not be a practical and inclusive assumption. The lack of trust can lead to reduced or dishonest contributions, in turn lowering the quality of data. The dependency on intermediaries can also introduce vulnerabilities and privacy risks. It can lead to a lack of transparency and accountability in the data generation pipeline.

This work advances the state-of-the-art in contributor-centric and decentralised synthetic data generation approaches. Building upon~\cite{oxford}, we develop a system that runs differentially private synthetic data generation algorithms in a scalable, decentralised and private manner, while the contributors of original data retain control of their data and their decision to participate in a given synthetic data generation process. Our approach requires no alteration in the synthetic data generation algorithm and thus, provides the same level of accuracy and differential-privacy guarantees as the approaches that assume a trusted centre. Unlike local differential privacy that has worse accuracy-privacy trade-off~\cite{openmined-localvsglobal}, our approach implements global differential privacy in a decentralised manner.

At the heart of this solution lies Solid (derived from social linked data), a decentralised Web specification pioneered by~\cite{sambra_solid_2016}. Solid lets people store their data securely in decentralised data stores called Pods and lets them control which application have access to their data. A variety of user data can be stored in Solid pods (further discussion on Solid in Section 3). Individuals may also be willing to contribute the data in their Pods to generate synthetic data for a variety of causes. However, for synthetic data generation, computations have to be performed over the data of multiple people. This is challenging without requiring a trusted centre. The challenge is further complicated by the fact that Solid Pods lack any local computation capability. \cite{oxford} proposed a modular architecture for integrating Secure Multi-Party Computation (MPC) with Solid, enabling arbitrary computations to be performed in a decentralised manner over data stored in Solid Pods. We build upon this recent research and show how differentially-private synthetic data can be generated in a scalable manner without compromising the autonomy and privacy of contributors. Instead of only relying on MPC, we offer a more scalable alternative that uses both MPC and a secure enclave (Intel SGX). We show how to implement different steps of synthetic data generation algorithms in MPC and SGX, utilising MPC where it offers the most value and using SGX elsewhere to overcome MPC’s performance challenges. We also analyse  various strengths and limitations of the approach.

\section{Problem Description}
	Due to page limit, we provide detailed background about personal data stores, Solid, multi-party computation, differential privacy and synthetic data generation in the supplementary material. We encourage the reader to consult the supplementary material if these concepts are unfamiliar.\footnote{Supplementary material is available at \url{https://goelnaman.github.io/upload/doc/papers/2025/ramesh_wiiat25_supp.pdf}}.
 	
	For the scope of this paper, a synthetic dataset is a substitute for an original dataset that has the same format and reflects the statistical properties of the original dataset, without reproducing the records in the original. In our settings, different individual data contributors hold their own data records in a decentralised manner. We are interested in generating synthetic data that mimics the properties of a dataset containing records of all individuals who are willing to participate in the synthetic data generation process, without requiring a trusted centre and ensuring that individuals have full autonomy and privacy. During the synthetic data generation process, the original data records should be kept confidential with the respective individuals and not revealed to others (i.e. input privacy). From the generated synthetic data, an adversary should not be able to make undesired inferences about the original data records or the individuals (i.e. output privacy).
	
	In this paper, we consider synthetic data generation algorithms that provide differential-privacy guarantees~\cite{dwork2006calibrating} for output privacy. However, it should be noted that the algorithms to generate differentially-private synthetic data, by themselves, provide neither decentralisation nor input privacy. We therefore need additional mechanisms. For decentralisation, we need decentralised storage, access control and reduced dependence on a trusted centre for the computation steps in the synthetic data generation algorithms. For input privacy, we need cryptographic mechanisms for private computation on input data.
	
	\cite{oxford} proposed a novel architecture, Libertas, to integrate Personal Data Stores with Secure Multi-Party Computation (MPC). Personal Data Stores provide individuals decentralised storage and access control for their data records. Secure Multi-Party Computation (MPC) provides input privacy and allows individuals to participate in arbitrary computations over their data records collectively with- out relying on a trusted centre. Decentralised synthetic data generation with input and output privacy is thus one example of a use-case that can be realised using Libertas. However, the scalability of this approach remains a challenging problem due to high computation and communication costs of the MPC protocols. We address this technical challenge in our work.
	
	\section{Proposed Approach for Scalable Decentralised Synthetic Data Generation}
	
	\begin{figure}[htbp]
		\includegraphics[width=1\linewidth]{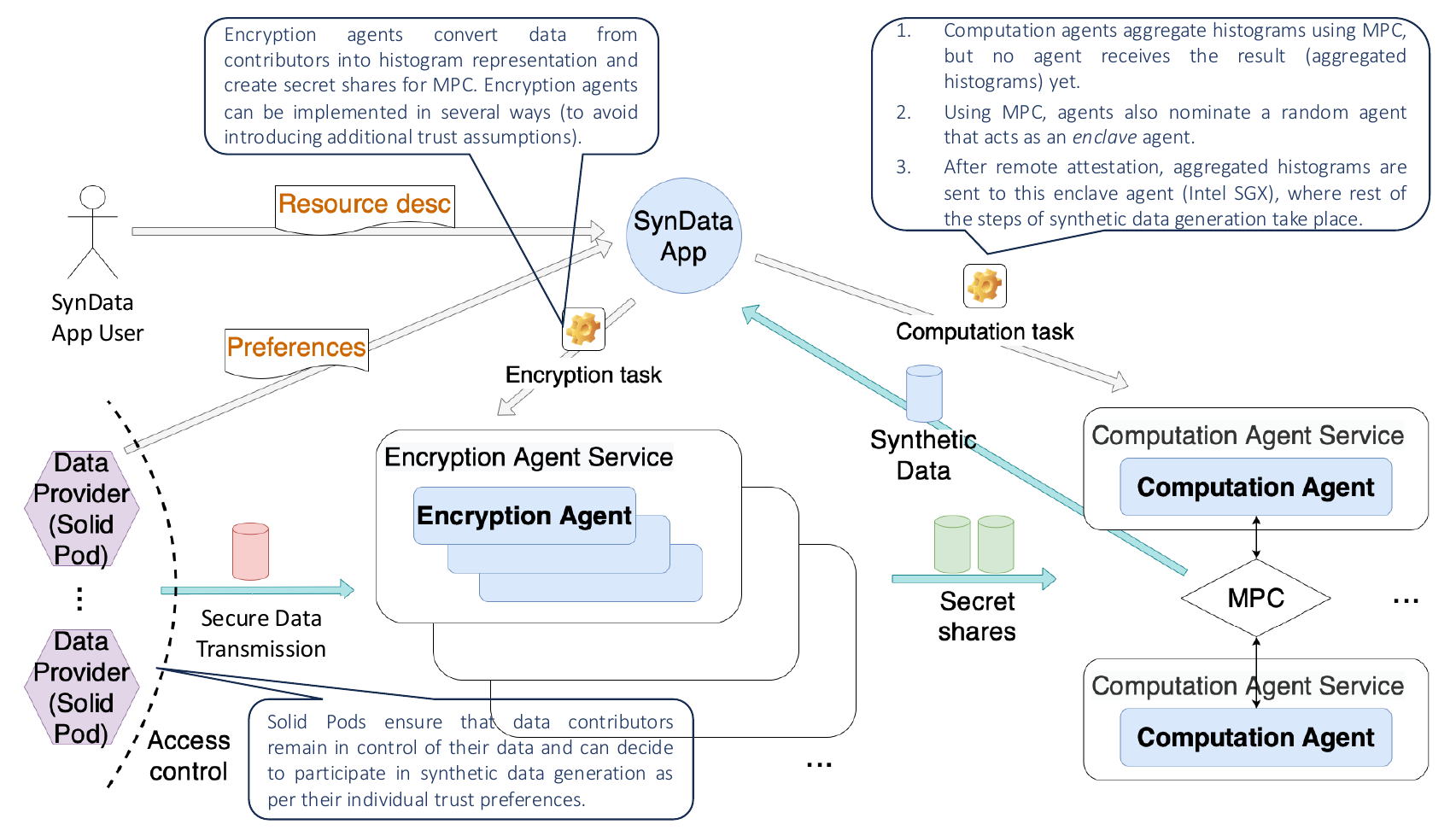}
		\caption{For generating differentially-private synthetic data from personal data stored in Solid pods, we adapt the Libertas architecture such that MPC is used for histogram aggregation and nominating a random enclave agent only. Subsequent steps of synthetic data generation are executed in the enclave (Intel SGX) after remote attestation.}
		\label{fig:mpc_sgx}
	\end{figure}
	
	To address the issue of high computation and communication costs incurred in Libertas due to its dependence on Secure Multi-Party Computation, we propose an adaptation tailored for differentially-private synthetic data generation algorithms. Instead of only relying on MPC, we propose a more scalable alternative that uses both MPC and secure enclaves. A secure enclave is a secure area in the main processor of a device; examples include Intel SGX~\cite{costan2016intel} or AMD SEV~\cite{sev2020strengthening}. We implement different steps of synthetic data generation algorithms in MPC and SGX, utilising MPC where it offers the most value and using SGX in other parts of the pipeline to overcome the performance challenges.
	
	\subsection{Steps}
	\begin{enumerate}
		\item Anonymisation: In the first step, data providers can remove personally identifiable information (PII) from data records. As users of Solid can set access control to each individual data item in their records, this step can be trivially performed by the data provider through the access control mechanism.  If desired, this step can also be enforced at app level using structured queries to ignore certain RDF vocabularies that are known to contain PII such as \verb|schema:identifier| \cite{murthy2019comparative}.
		
		\item Client-Side Aggregation: The next step is for the chosen encryption agents in the Libertas architecture to run the client code~\footnote{The Libertas architecture allows encryption agents to be either implemented by independent service providers from which individuals can select their respective encryption agent based on their trust preferences or implemented locally for e.g. on the device of individual, web browser or implemented by the Solid server that hosts the Pod of individual. This provides flexibility in implementation depending on the requirements, while making explicit the trust assumptions introduced under different situations.}. This involves reading the access authorised data from the Pod of participating data provider and converting it into a histogram representation, also performing any histogram binning along the way. This setup allows for settings where each data provider manages more than one data record.
		
		\item MPC-Based Histogram Aggregation: The third step is for the computation agents in the Libertas architecture to run the MPC code. This involves reading shares of the histogram representations from the clients and computing the aggregate histograms through an addition of arrays (or matrices) operation. No agent receives the result (aggregate histograms) yet.
		
		\item Differentially-Private Synthetic Data Generation: The next step is to execute the remaining steps of differentially-private synthetic data generation algorithm. We do not perform this step in MPC; more details about this step as in the following text.
	\end{enumerate}
	
	\subsection{Separating Noise Addition and Generation from MPC}\label{downstream}
	
	We propose that by separating the preliminary (histogram aggregation) step in MPC from the subsequent steps, we can improve the scalability of privacy friendly synthetic data generation in decentralised contexts. For example, depending on the synthetic data generation algorithm, these subsequent steps may involve adding noise to the histograms using the appropriate mechanism based on the designated privacy budget to ensure the measurements are differentially-private and then run a generation algorithm such as PGM on the measurements that can be released to the MPC app user. Similarly, in the case of MWEM synthetic data generation algorithm~\cite{hardt2012simple}, this involves running the iterative algorithm where measure and generate steps are more interweaved. Note that we do not modify the steps of the algorithm or propose a new algorithm. The aggregate histograms that are needed as input in the synthetic data generation algorithm are computed from the individual data points using MPC. This aggregated histogram is then used by the algorithm but the algorithm itself is not implemented in MPC (unlike closely related work of \cite{oxford,pereira2022secure}). We therefore inherit the differential- privacy guarantees from the respective algorithm for the final output. Let us now address an intermediate vulnerability that is introduced as a result of this separation from MPC and how we alleviate it.
	
	In particular, between steps 3) and 4), we need to trust a computation agent with the aggregated histogram, who then runs a differentially-private synthetic data algorithm. If this agent is compromised, an adversary can access non-private aggregated histogram.
	
	What the adversary could learn about individuals participating in the data analysis depends on the nature of the dataset. For example, in non-homogeneous populations, identifiable outliers in the histogram could compromise the privacy of participants from certain subgroups. This is of particular concern if it involves protected characteristics or individuals from marginalised groups. To mitigate against these attacks, we use the following provisions to strengthen the privacy assurance of our approach.
	
	\subsubsection{Random Selection}
	In a naive implementation of MPC, all parties receive the same output, i.e. the result of evaluating the arithmetic circuit. However, it is possible for different parties to receive different outputs or no output at all. We delegate the subsequent tasks to a randomly chosen computation party instead, i.e. only the chosen party receives the aggregated histogram output. The random selection is part of the MPC circuit as a joint random number generation procedure and hence adherence is enforced by the MPC protocol used (and with the underlying security parameters).
	
	Random selection reduces the chance of a malicious party getting access to the aggregated histogram. Furthermore, it is impossible for the adversary to cheat the random number generation through its choice of inputs alone. It is worth recalling that MPC still protects the input data (individual records) as long as the protocol security assumptions hold, limiting the attack surface.
	
	\subsubsection{Secure Enclave}
	A secure enclave is a secure area in the main processor of a device; examples include Intel SGX~\cite{costan2016intel} or AMD SEV~\cite{sev2020strengthening}. With appropriate complementary mechanisms, secure enclaves are a promising approach to maintaining the confidentiality and integrity of code and data located inside them, without relying on a trusted operating system. Therefore, many cloud service providers such as Azure offer enclaves based “confidential computing” feature for sensitive use-cases. The MPC histogram generation proceeds as before. Additionally, a computation agent is nominated to serve as the enclave agent. The enclave agent can be nominated using the MPC based random selection as discussed above.
	Before dispatching the code for the remaining synthetic data generation steps to the enclave agent, the encryption party can verify the identity of the enclave through remote attestation. Remote attestation, which allows a remote party to verify the contents of the program that is running in the enclave with a certificate generated by the underlying hardware. After the MPC has concluded, the output (aggregated non-private histogram) is sent to the enclave via a secure channel for executing rest of the steps of differentially-private synthetic data generation process. Typically an extension of SSL/TLS which forces the enclave endpoint to incorporate its attestation proof with its certificate serves as the channel~\cite{knauth2018integrating}.
	We provide an overview of our approach in Figure~\ref{fig:mpc_sgx}, an adapted version of the one presented in~\cite{oxford}.
	
	\subsection{Discussion} 
	We note that SGX is vulnerable to side channel attacks~\cite{nilsson2020survey} and the use of SGX in our proposal introduces an implicit trust assumption on hardware manufacturer. However, since the first step in our solution uses MPC and enclave agent is also chosen using MPC, we not only keep the solution decentralised but the risks due to SGX’s vulnerabilities are also significantly reduced.  Thus, in order to utilise secure enclave within the Libertas architecture in a principled manner, we separate the computation steps in the algorithm into steps that should be executed in MPC and secure enclave while balancing the privacy, decentralisation and scalability trade-offs.
	
	We also note that while setting up an enclave, all the dependencies must be loaded into the enclave file system ahead of its creation. This limits the libraries that the code running in the enclave can access. Thus for convenience, the services offered by the enclave agent is baked into the architecture, at the cost of versatility. 
	
	\subsection{Assumptions} The proposed approach inherits threat model assumptions from parent technologies and architectures including Solid~\cite{sambra_solid_2016,mansour2016demonstration}, MPC~\cite{evans2018pragmatic}, the MPC framework used in our implementation: MP-SPDZ~\cite{keller2020mp}, Libertas~\cite{oxford} , differential-privacy~\cite{dwork2006calibrating} and Intel SGX~\cite{nilsson2020survey}. The assumptions include but are not limited to, encryption agents and computation agents in the Libertas architecture behaving as specified in the trust preferences of the data providers, and appropriate MPC protocol being used for different settings (for e.g. honest majority or dishonest majority setting).
	
	\section{Empirical Evaluation}\label{sec:evaluation}
	We implemented the above approach and performed comprehensive experiments to understand the strengths and limitations of the proposed approach. In this section, we will first discuss implementation related details, following by experimental setup and the results.
	
	\subsection{Implementation Details}
	Apart from the Solid development framework and the Libertas implementation~\cite{oxford}, we use the following frameworks and libraries in our implementation.~\footnote{Source code is available at \url{https://github.com/OxfordHCC/libertas}}
	
	\subsubsection{MP-SPDZ}
	MP-SPDZ is a framework for benchmarking multi-party computation tasks. MP-SPDZ uses a syntax similar to Python for writing high level program that is compiled to a low level byte code executed on a virtual machine \cite{keller2020mp}. Each protocol is implemented by a different virtual machine. Similar to \cite{oxford}, we use MP-SPDZ in our experiments because it enables easy testing of different protocols from a single high level program and for its comprehensive metrics.
	
	\cite{oxford} implemented the encryption agents using the client mechanism of the MP-SPDZ framework for secret sharing with players, which in turn is based on the SPDZ protocol \cite{damgaard2016confidential}. We follow the same and we also use it for the enclave agent. Although it does not send (unlike encryption agents) any input to the computation parties, we use the \textit{client} mechanism of MP-SPDZ to allow it to receive the aggregated histogram.
	
	\subsubsection{Gramine}
	Gramine is a library OS that is used to run unmodified Linux applications (e.g. the Python interpreter) in an SGX enclave \cite{tsai2017graphene}. While the Intel SGX SDK exposes a low-level C/C++ interface for writing enclave applications, we use Gramine for portability and faster development cycles. Gramine works much like a unikernel by exposing a set of system call implementations as a user space library, serving as a compatibility layer. We provide a manifest file with the associated source code to configure the application environment and isolation policies.
	
	\subsubsection{Marginal-Based Inference (MBI)}
	The Marginal-Based Inference (MBI) library exposes procedures that take as input noisy measurements (e.g. differentially private marginals) and generates synthetic data~\cite{private-pgm}. While selection of the right queries is an important question \cite{mckenna2022aim}, we restrict attention to the scalability and MBI provides a generic interface with a range of implemented algorithms for comparison. As with MP-SPDZ, we use MBI for easy testing of different generation algorithms and comparability of benchmarking results.
	
	\subsection{Experimental Setting}
	\subsubsection{Setup}
	Our computational and encryption agents were deployed on a single server connected over a virtual LAN. This was to focus on the computational cost of the approaches while controlling for network factors such as latency and bandwidth that would affect the scalability. We acknowledge that these are important as the agents may be connected over a WAN, hence we report the size of data transfers in each computation.
	
	We deploy 3 computation agent servers and 2 encryption agent servers, unless stated otherwise. This is because for honest majority protocols, 3 is the minimum number of players. Furthermore, \cite{oxford} show that Libertas with fewer MPC players scale better. Hence, any improvement our approach offers in this setting should also translate to cases with greater number of players.
	
	\subsubsection{Platform}
	The experiments were conducted on an Azure DV4 VM running on an Intel Xeon Platinum 8370C (2.8GHz) with 4 vCPUs and 32 GB RAM. This architecture includes support for the SGX instruction set and this VM series comes with SGX enabled in the BIOS. A distribution with Linux kernel of version at least 5.11 is used as it includes support for SGX drivers (we use Ubuntu 22.04).
	
	We increased the maximum number of open files to 65536 as we otherwise quickly run out of file descriptor resources during our experiments. Nevertheless, in certain (e.g. dishonest majority) settings we encounter resource limitations and hence must limit the scope of our experiment to fewer data providers.
	
	\subsubsection{Benchmark Datasets}
	For evaluating the proposed approach, we simulate different settings in which data providers have data records stored in their Solid pods. For the first part of our experiments, we use 1-dimensional simulated data records. Please note the distinction between `simulated data' and `synthetic data'. In our paper, synthetic data is the data produced by a differentially-private generation algorithm that takes real data from data providers as input. In the first part of our experiments, we simulate this real data using `simulated data', which would then be the input for synthetic data generation algorithm. The simulated data was uniformly sampled from a range of 0 to 20. There are two scenarios that we consider here. The first is the fixed total data setting where we fix the total number of data records at 10000 and evenly distribute it among the data providers. The second is the variable total data setting where we fix the number of data records per data provider at 100. This latter situation gives us between 1000 and 100000 data records (the number of data providers range from 10 to 1000).
	
	For the second part of experiments, where the performance of marginal-based generation algorithms under our approach is benchmarked, we use real-world datasets that are common benchmarks in the synthetic data generation literature. In particular, we rely on the Adult \cite{kohavi1996scaling} and the Titanic \cite{encyclopediatitanica} datasets. There are a few preprocessing steps that we undertake before storing the data records in Solid Pods. Since many generation algorithms (including the MWEM algorithm) assume discrete-valued data, we remove certain continuous values attributes and discretise others. We further remove data records with missing values.
	
	\begin{table}[h!]
		\centering
		\caption{A summary of the real-world datasets used in the experiments (after preprocessing). Total domain size was rounded to one significant figure.}
		\begin{tabular}{ | m{2.8em} | m{0.9cm} | m{1.3cm} | m{1.3cm} | m{1.7cm} | } 
			\hline
			\textbf{Dataset} & \textbf{Records} & \textbf{Dimensions} & \textbf{Min-Max Domains} & \textbf{Total Domain Size}  \\
			\hline
			Adult & 48842 & 14 & 2-100 & $4 \times 10^{17}$  \\
			\hline
			Titanic & 713 & 7 & 2-90 & $2 \times 10^5$  \\
			\hline
		\end{tabular}
		\label{tab:tab10}
		
	\end{table}
	
	Table \ref{tab:tab10} shows a summary of the datasets after preprocessing. Both real-world datasets have much larger domains than the simulated data.
	
	Since the number of instances in Adult (48842) is much larger than the maximum number of data providers we consider (1000), we distribute data instances to data providers in two different ways, much akin to the difference between the variable total and fixed total datasets described before. In the first, each data provider receives a single instance (e.g. selected randomly without replacement) from the Adult dataset. This is more in line with the traditional thinking of a Pod as being controlled by and holding a single individual's data. In the second, all the records from Adult are distributed roughly equally among the different data providers.
	
	\subsubsection{Client Binning}
	We use a pre-specified number of bins in histogram for inducing a manageable dimensionality. This conversion of individual data points to the bin format can be done either in the encryption server via client code or in the computation server via MPC code. The binning strategy is pre-agreed through client code. Results from \cite{oxford} find that performing binning in the client code results in a significant performance gain. We use this so-called \textit{client-binning} optimisation in all our experiments involving the MWEM synthetic data generation algorithm.
	
	\subsection{Reported Metrics}\label{sec:metrics}
	
	\begin{enumerate}
		\item \textbf{Time:} The total time taken to complete a given MPC computation, including any setup time, for e.g., establishing connections, fetching data, shares of data being received from the encryption server clients etc.
		While reporting this metric for our approach, we add the time for performing the measure and generate steps in SGX.
		\item \textbf{Communication Rounds:} The number of rounds of communication required for the players to finish a given computation. This differs from the number of communications which is the number of rounds times the number of players (for the protocols we consider, may differ in other protocols).
		\item \textbf{Local Data:} The amount of data being transmitted by a single player. Usually this value does not vary greatly between different players for the MPC protocols that we use. We report the local data measured at player 0 for consistency.
		\item \textbf{Global Data:} The total amount of data exchanged during all communications between players over the course of the MPC computation.
	\end{enumerate}
	
	Communication rounds, local and global data are standard metrics provided by the MP-SPDZ framework itself; we did not measure these metrics using a separate code or tool to ensure easy reproducibility.\\
	
\noindent \textbf{Remark regarding accuracy metrics:} Accuracy of synthetic data and its utility in downstream use-cases are very important considerations, but those aspects are influenced by the synthetic data generation algorithm and the inputs of the algorithm. In our paper, the focus is on the underlying computation system that executes synthetic data generation algorithm. We do not modify the algorithms themselves or the input to the algorithms; our research is concerned with the efficient implementation of these algorithms in a decentralised computing environment. Therefore, accuracy metrics are not relevant metrics in the context of this work. This is also verified by prior work (e.g. by ~\cite{pereira2022secure}) which empirically showed that the accuracy and downstream utility of the synthetic data does not change merely because the generation algorithm is implemented in a different computing environment (as long as the algorithm, the parameters of the algorithm and the input real data used for synthetic data generation remain the same). We encourage readers who are curious about the accuracy and utility of synthetic data in general, to refer to the research papers about respective synthetic data generation algorithms(e.g.~\cite{hardt2012simple,mckenna2019graphical}) or the papers about use-cases that employ synthetic data in downstream applications(e.g. ~\cite{mckenna2021winning}). Such discussion on accuracy is not included in our paper in order to avoid confusion and to keep the focus on metrics that are directly relevant to our contributions.
	
	\subsection{Empirical Analysis}
	We now present results from experiments, comparing our proposed hybrid approach using MPC and SGX versus MPC only, for the MWEM synthetic data generation algorithm~\cite{hardt2012simple} across various simulated datasets and MPC protocols with different adversarial assumptions. The MPC only implementation is taken from Libertas~\cite{oxford}, while the MPC and SGX implementation is ours.
	
	We first discuss the results obtained on the simulated data and the MASCOT MPC protocol\cite{keller2016mascot}. We run the MASCOT protocol to simulate a dishonest majority setting. 
	
	\begin{figure}[H]
		\centering
		\begin{subfigure}[b]{0.475\linewidth}
			\centering
			\includegraphics[width=\textwidth]{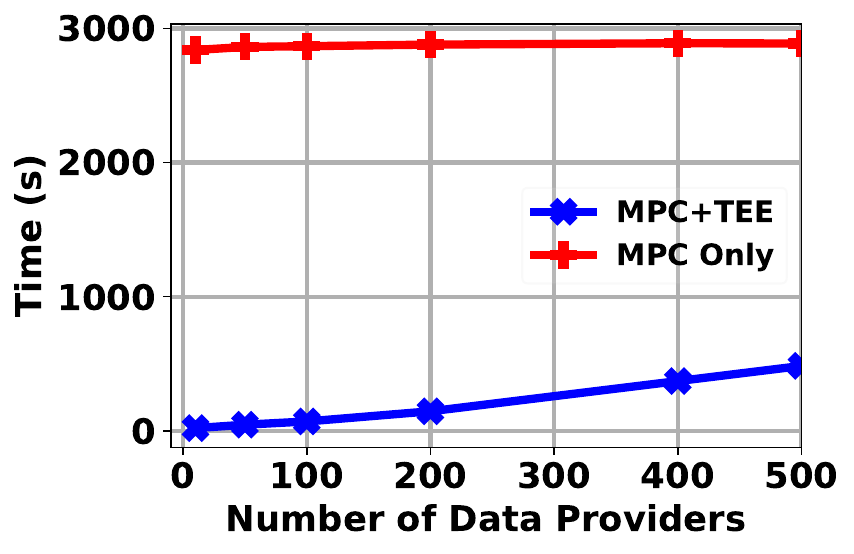}
			\caption[]%
			{{\small Time}}
		\end{subfigure}
		\hfill
		\begin{subfigure}[b]{0.475\linewidth}  
			\centering 
			\includegraphics[width=\textwidth]{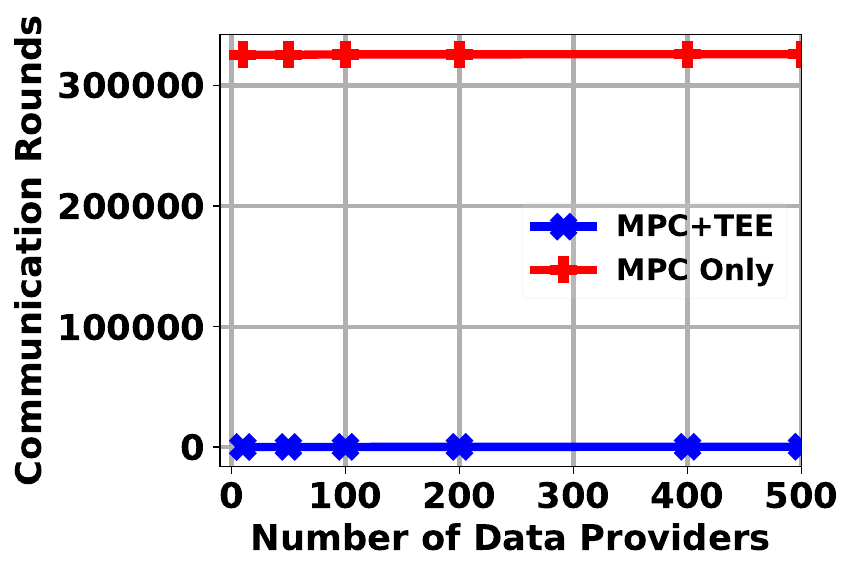}
			\caption[]%
			{{\small Communication Rounds}}    
		\end{subfigure}
		\vskip\baselineskip
		\begin{subfigure}[b]{0.475\linewidth}   
			\centering 
			\includegraphics[width=\textwidth]{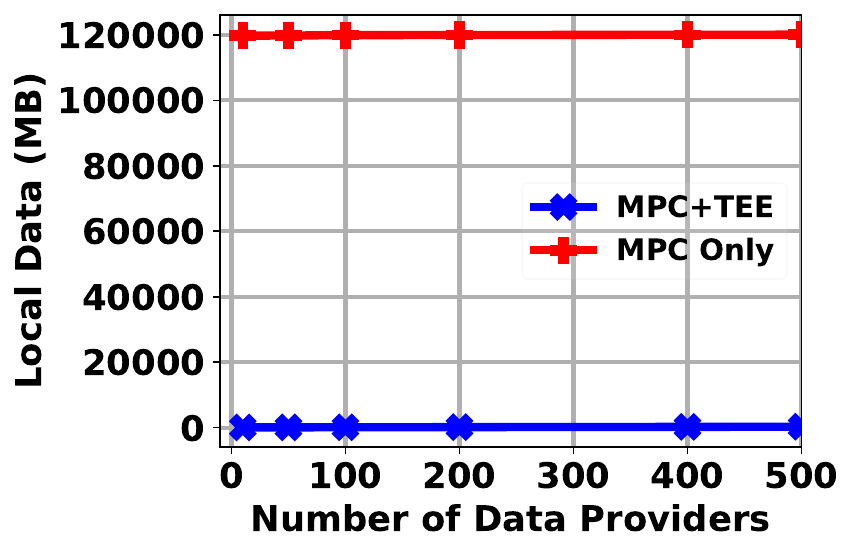}
			\caption[]%
			{{\small Local Data}}    
		\end{subfigure}
		\hfill
		\begin{subfigure}[b]{0.475\linewidth}   
			\centering 
			\includegraphics[width=\textwidth]{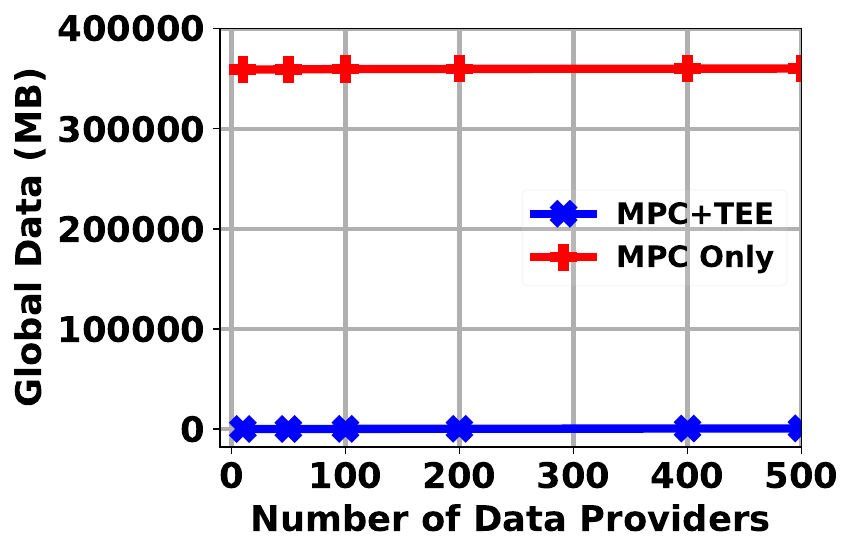}
			\caption[]%
			{{\small Global Data}}    
		\end{subfigure}
		\caption[]
		{\small Comparison of MPC Only and MPC+SGX Approaches [MWEM; MASCOT protocol, fixed total data (simulated) i.e. 10000 data points divided equally among data providers; 10 bins, $\epsilon = 2$, $T=30$.]} 
		\label{fig:fixed_artificial}
	\end{figure}
	
	Figure~\ref{fig:fixed_artificial} shows the results for the setting in which we fix the total number of data points among all data providers at 10000 and distribute them equally among them. MWEM algorithm is run with the client-binning optimisation using a fixed number of bins (10), epsilon value 2 ($\epsilon = 2$), and 30 iterations ($T = 30$). The reported time includes the time in communication setup. We observe that running MWEM algorithm with our approach takes less than 500 seconds, while MWEM with MPC only takes almost 3000 seconds even for just 10 data providers. While the computation time in both settings grows linearly with data providers, the faster growth in the MPC+SGX setting can be attributed to initial setup time where shares are received from the encryption agents, which increases with more data providers and may dominate a smaller computation time. \cite{oxford} also discuss computation time and setup times separately and show that computation time indeed grows much slowly than the total time (computation + setup time). 
	We also observed that the behaviour w.r.t. local and global data may vary with the implementation of the MPC program but generally, grows slowly with the number of data providers. We zoom into the graphs by showing one observation point in Table \ref{tab:tab1}. In the MPC only setting, the amount of global data exchanged to complete the computation is nearly 360 \textbf{GB}, compared to the much more manageable size of 357 \textbf{MB} in our approach. This is an orders of magnitude improvement.
	\begin{table}[h]
		\centering
		\caption{MASCOT with 10000 data items divided among 100 data providers. Time is rounded to 0.01; $\pm$ indicates standard deviation. MWEM on fixed total dataset (simulated) with 10 bins, $\epsilon = 2$, $T=30$.}
		\begin{tabular}{ | m{4.7em} | m{1.8cm} | m{1.0cm} | m{1.2cm} | m{1.2cm} | } 
			\hline
			\textbf{Approach} & \textbf{Time(s)} & \textbf{Rounds} & \textbf{Local Data(MB)} & \textbf{Global Data(MB)} \\
			\hline
			MPC Only & $2860.68 \pm 9.29$ & 325796 & 119936 & 359514 \\
			\hline
			MPC+SGX & $73.65 \pm 14.96$ & 208 & 118.98 & 356.87 \\
			\hline
		\end{tabular}
		\label{tab:tab1}
	\end{table}
	
	Figure~\ref{fig:variable_artificial} in Appendix~\ref{appendix:first} shows the corresponding results when we fix the total number of data points per data providers at 100. Figure~\ref{fig:fixed_artificial} and ~\ref{fig:variable_artificial} show very similar results due to the fact that the number of data points contributed by data provider does not matter as much in our implementation because the client code converts raw data from data providers into histogram representation and the rest of the process only deals with the histogram representation. Therefore, the additional cost is mostly limited to the process of converting raw data into histogram representations and further dominating costs are nearly independent of the number of data points.
	
	Appendix~\ref{appendix:second} analyses the effect of number of iterations (a hyper-parameter of the synthetic data generation algorithm) on the time as well as error.
	
	Due to page limit, so far we discussed results with MASCOT MPC protocol (dishonest majority setting). Appendix~\ref{appendix:third} further shows the results Shamir MPC protocol (honest majority setting), showing that effectiveness of our approach in both settings.

	\subsection{Benchmarks on Real Datasets}
	
	Having carefully analysed the difference between the MPC only and the MPC+SGX approaches using simulated datasets, we now benchmark the MPC+ SGX approach using real-world datasets. In particular, we benchmark the performance of different synthetic data generation algorithms taken from the MBI (Marginal-Based Inference) library~\cite{private-pgm}. In particular, we benchmark PGM (\texttt{mbi.FactoredInference}) and Local Consistency (\texttt{mbi.LocalInference}) algorithms in the proposed MPC+SGX approach with common datasets like Adults and Titanic. In the figures in this section, the results of the MPC+SGX approach will also be more clearly visible since we will not include the MPC only approach which would increase the scale of Y-axis in the the figures.
	
\begin{figure}[H]
		\centering
		\begin{subfigure}[b]{0.475\linewidth}
			\centering
			\includegraphics[width=\textwidth]{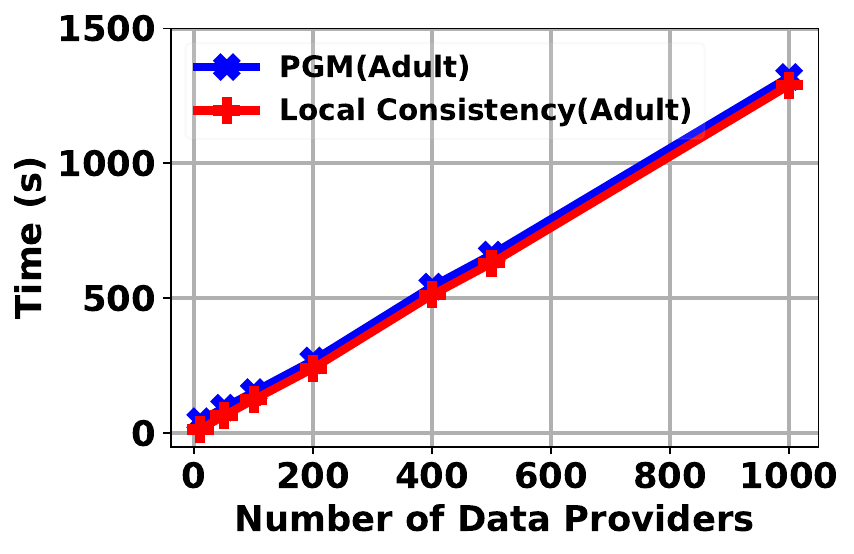}
			\caption[]%
			{{\small Time}}    
			\label{adult_time}
		\end{subfigure}
		\hfill
		\begin{subfigure}[b]{0.475\linewidth}  
			\centering 
			\includegraphics[width=\textwidth]{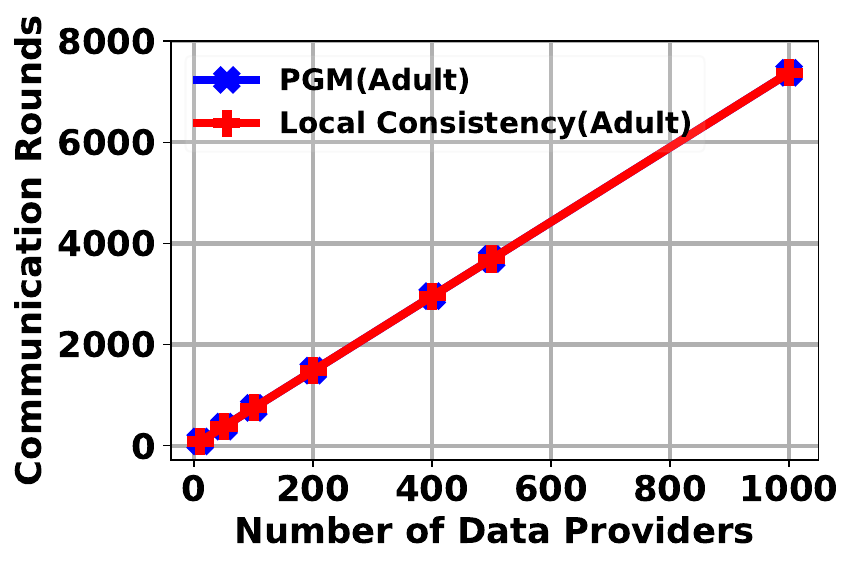}
			\caption[]%
			{{\small Communication Rounds}}    
		\end{subfigure}
		\vskip\baselineskip
		\begin{subfigure}[b]{0.475\linewidth}   
			\centering 
			\includegraphics[width=\textwidth]{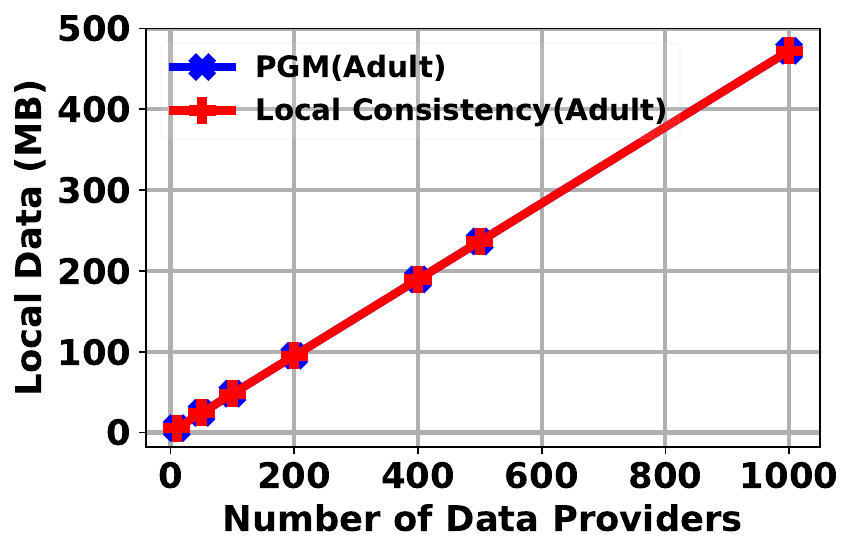}
			\caption[]%
			{{\small Local Data}}    
		\end{subfigure}
		\hfill
		\begin{subfigure}[b]{0.475\linewidth}   
			\centering 
			\includegraphics[width=\textwidth]{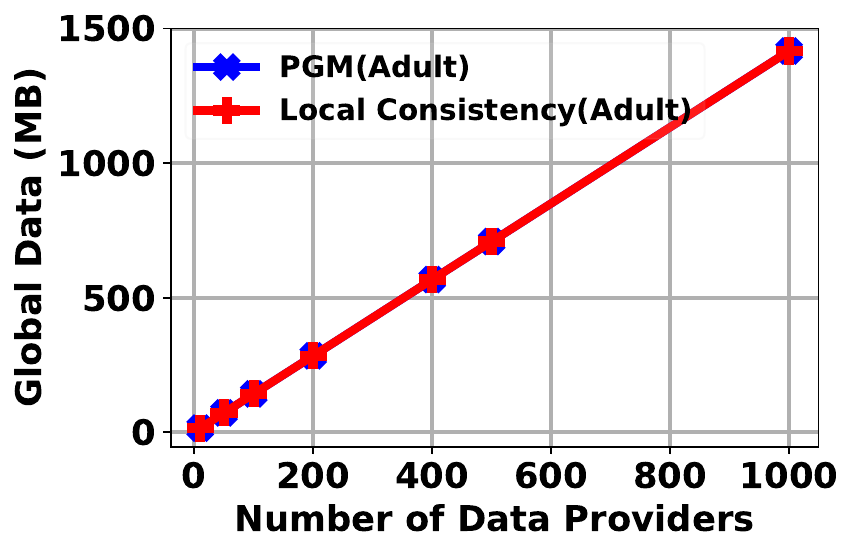}
			\caption[]%
			{{\small Global Data}}    
		\end{subfigure}
		\caption[]
		{\small Performance of the MPC+SGX approach using PGM and Local Consistency generation algorithms (30 iterations); Adult dataset; SHAMIR protocol, one data point per provider.]} 
		\label{fig:adult}
	\end{figure}

	We first benchmark the performance of the MPC+SGX approach with the PGM algorithm on 10 two-dimensional marginals of the Adult dataset using the Shamir protocol. The reason we did not consider a larger number of marginals is because of memory limitations we encountered on our compute while running PGM rather than the scalability of the approach to real world high-dimensional datasets. Each data provider represented a single record from Adults. PGM was run for 30 iterations with no change to the other hyper-parameters. The results are shown in Figure~\ref{fig:adult}. Despite the large number of attributes (14), many of which have a large domain (maximum 100), our approach keeps the computation time at less than 1400 seconds even with 1000 data providers. While amount of data exchanged grows linearly, it remains at a manageable 1400 MB global data for 1000 providers.
	
	Appendix D~\ref{appendix:forth} also shows the results of benchmarking with PGM algorithm on Titanic dataset.
	
	\section{Differentiation from Other Related Work}\label{sec:other_related}
	 We note two lines of related work and explain the difference from our approach. The first is local differential privacy methods~\cite{zhao2020local,cormode2018privacy,wang2019collecting,qin2017generating,ren2018textsf}. The accuracy of query results in the local model is typically orders of magnitude lower for the same privacy cost and the same query, compared to central differential privacy~\cite{dwork2014algorithmic}. On the other hand, we use central differential privacy using other privacy-enhancing technologies such as MPC and SGX and therefore, the privacy-utility trade-off is the same as in central model. We show that one can limit MPC use to strictly necessary for addressing the scalability challenge. The second line of work either proposes to use blockchains for decentralised data and access control or focuses on synthetic data generation process or other computations from distributed data. Examples from this very large body of work include~\cite{zyskind2015enigma,zyskind2015decentralizing,golob2023decentralized,pereira2022secure,veeraragavan2023securing,hynes2018demonstration}. However, cost and privacy concerns make on-chain data storage of personal data infeasible, and off-chain storage which is often suggested as an alternative to address this concern, does not really solve the problem but merely delegates it. Further, the guarantees provided by blockchains such as immutability are not required (and may even be undesired) in many use-cases and come with significant costs. On the other hand, we use Solid (a personal data store) which provides the necessary features such as decentralised storage, efficiency and access control, and is based on open web protocols.
	
	\section{Conclusion}
	In this paper, we advance state-of-the-art in contributor-centric approach to responsible synthetic data generation. Personal data stores such as Solid provide individuals ultimate control over their data on the Web. Users keep their data in Pods (Personal Online Datastores). A Pod provides granular control to users over which applications can access which data as well as secure transmission of data for authorised requests. The apps and services that they use (for e.g. various web based applications, online social platforms, health service applications etc), also read and write data to the user's Pod under user-specified preferences. In the proposed approach for synthetic data generation, individuals use Solid’s access control mechanism to decide whether they want to participate in the synthetic data generation process (for e.g., by taking into account what is the purpose of synthetic data generation and which real data is required as input). We show how participating individuals can then generate differentially-private synthetic data with the help of secure multi-party computation (MPC) and Intel SGX. We show that the resulting system can be used to significantly improve the scalability of synthetic data generation by offering orders of magnitude of reduction in running time and communication overhead compared to the relevant state-of-the-art baseline. Such improvements are crucial for making privacy-enhancing technologies attractive and for increasing their adoption in the real-world.
	
	\section{Limitations and Future Work}
	In future work, it would be interesting to explore other personal data stores such as openPDS and Databox with different capabilities than Solid. We expect to see similar trends on those platforms under similar experiments, but there could be interesting variations in the cause of performance bottlenecks and worth investigating. Similarly there are other frameworks for MPC, many derived from MP-SPDZ and some independent ~\cite{duan2020introduction}. While communication costs impose a natural challenge, more efficient implementations of the underlying protocols could also contribute to more scalable synthetic data generation. Finally, we have focused on tabular data in this work whereas Solid also has support for unstructured data such as text and images. There has been much success in private synthetic data generation of images using large generative models. An interesting direction for future inquiry would be to improve their scalability in a decentralised setting.
    
\bibliographystyle{unsrturl}
\bibliography{references}

\appendices
\section{Fixed Total Number of Data Points Per Data Provider}\label{appendix:first}
	
\begin{figure}[!h]
	\centering
	\begin{subfigure}[b]{0.475\linewidth}
		\centering
		\includegraphics[width=\textwidth]{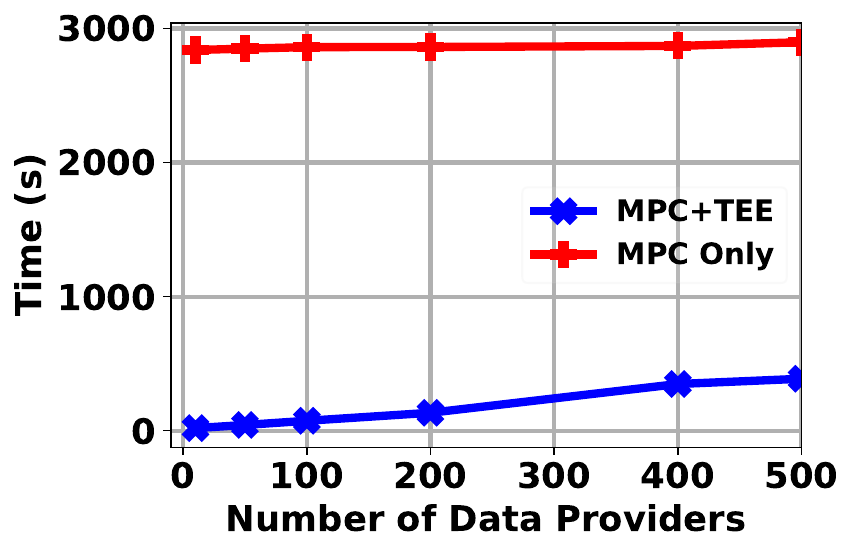}
		\caption[]%
		{{\small Time}}    
	\end{subfigure}
	\hfill
	\begin{subfigure}[b]{0.475\linewidth}  
		\centering 
		\includegraphics[width=\textwidth]{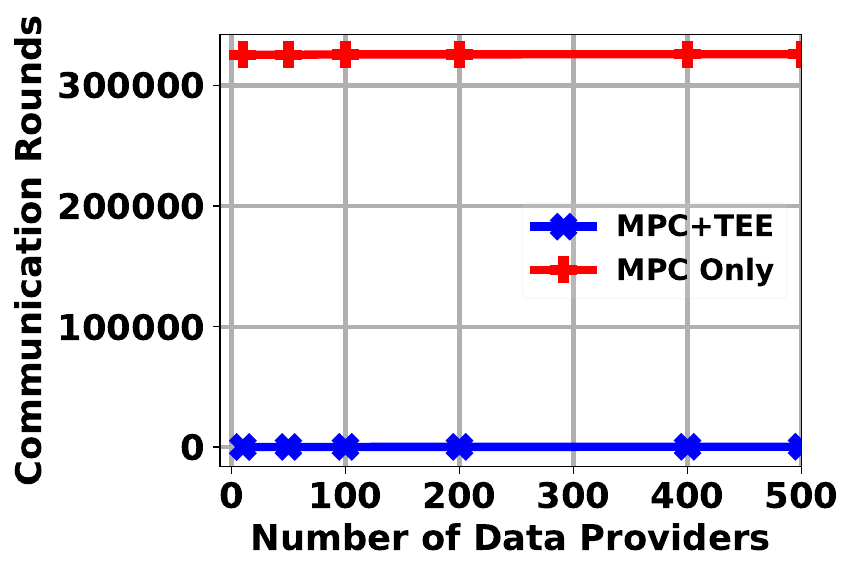}
		\caption[]%
		{{\small Communication Rounds}}    
	\end{subfigure}
	\vskip\baselineskip
	\begin{subfigure}[b]{0.475\linewidth}   
		\centering 
		\includegraphics[width=\textwidth]{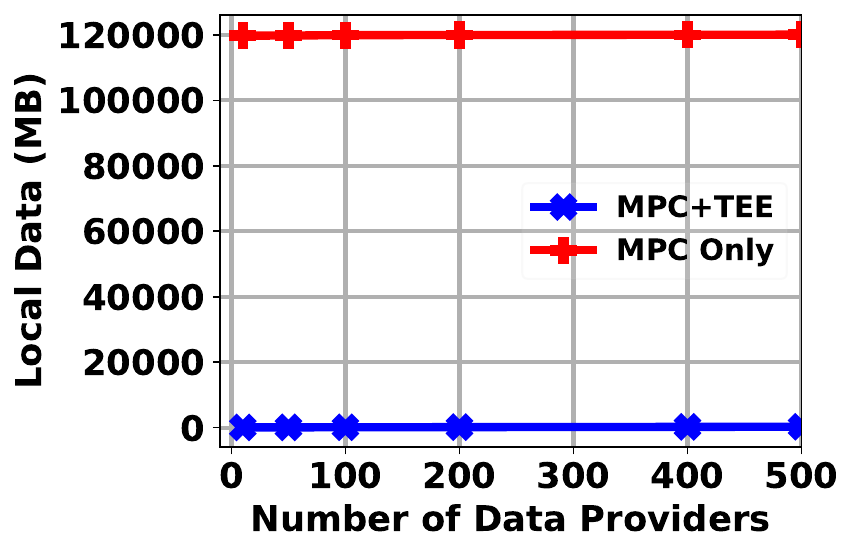}
		\caption[]%
		{{\small Local Data}}    
	\end{subfigure}
	\hfill
	\begin{subfigure}[b]{0.475\linewidth}   
		\centering 
		\includegraphics[width=\textwidth]{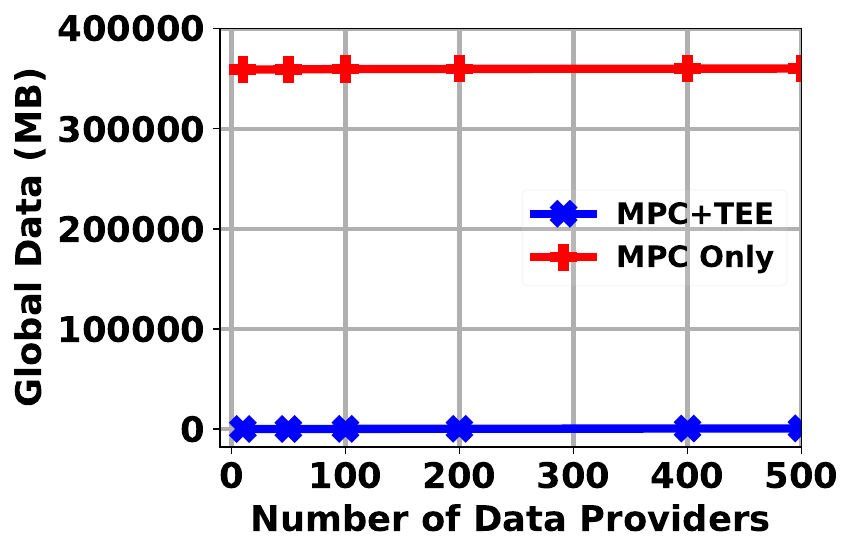}
		\caption[]%
		{{\small Global Data}}    
	\end{subfigure}
	\caption[]
	{\small Comparison of MPC Only and MPC+SGX Approaches [MWEM; MASCOT protocol, variable total data (simulated) i.e. 100 data points per provider; 10 bins, $\epsilon = 2$, $T=30$.]} 
	\label{fig:variable_artificial}
\end{figure}

\section{Discussion about MWEM parameters (number of iterations and number of bins)}\label{appendix:second}

\begin{figure}[!h]
 	\centering
 	\begin{subfigure}[b]{0.475\linewidth}
 		\centering
 		\includegraphics[width=\textwidth]{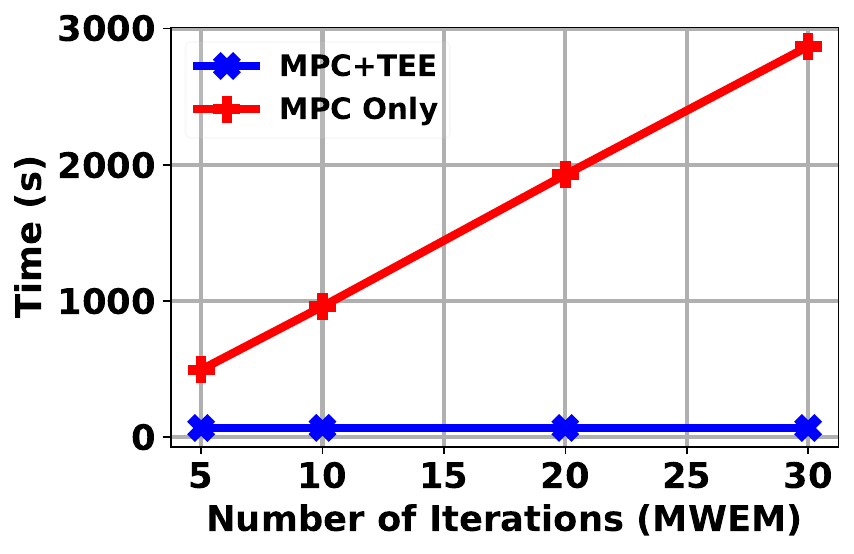}
 		\caption[]%
 		{{\small Fixed total data (simulated) i.e. 10000 data points divided among 100 data providers.}}    
 	\end{subfigure}
 	\hfill
 	\begin{subfigure}[b]{0.475\linewidth}  
 		\centering 
 		\includegraphics[width=\textwidth]{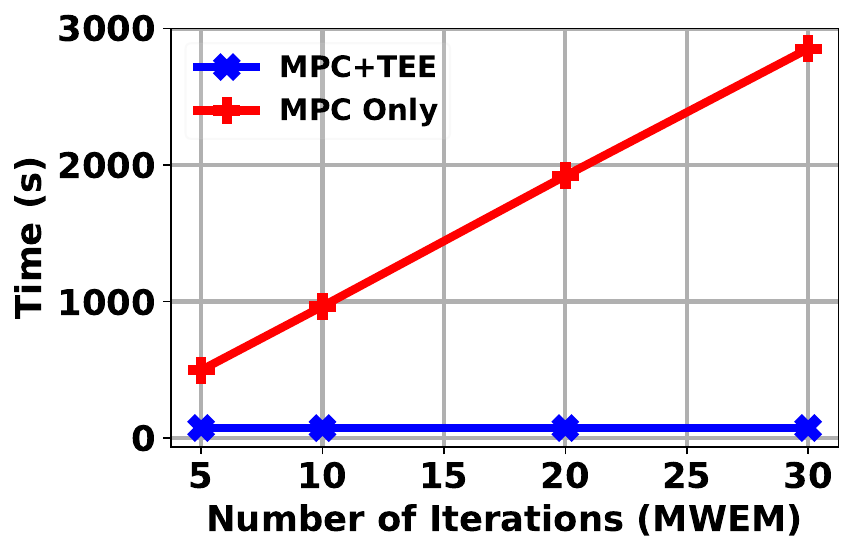}
 		\caption[]%
 		{{\small Variable Total Data (simulated) i.e. 100 data points per provider.}}    
 	\end{subfigure}
 	\caption[]
 	{\small Comparison of time in MPC Only and MPC+SGX Approaches, for different number of iterations of MWEM. MASCOT protocol, 100 data providers, 10 bins, $\epsilon = 2$, $T=100$.} 
 	\label{fig:time_iter}
 \end{figure}

In Figure~\ref{fig:time_iter} we show the time taken with the number of iterations in the MWEM algorithm, an important hyper-parameter in any iterative algorithm. Clearly, time increases at a much faster rate in the MPC only approach compared to the hybrid MPC+SGX approach. The reason for similarity in the fixed data and variable data setting is the same as discussed previously. 
 
\begin{figure}[H]
	\centering
	\begin{subfigure}[b]{0.475\linewidth}
		\centering
		\includegraphics[width=\textwidth]{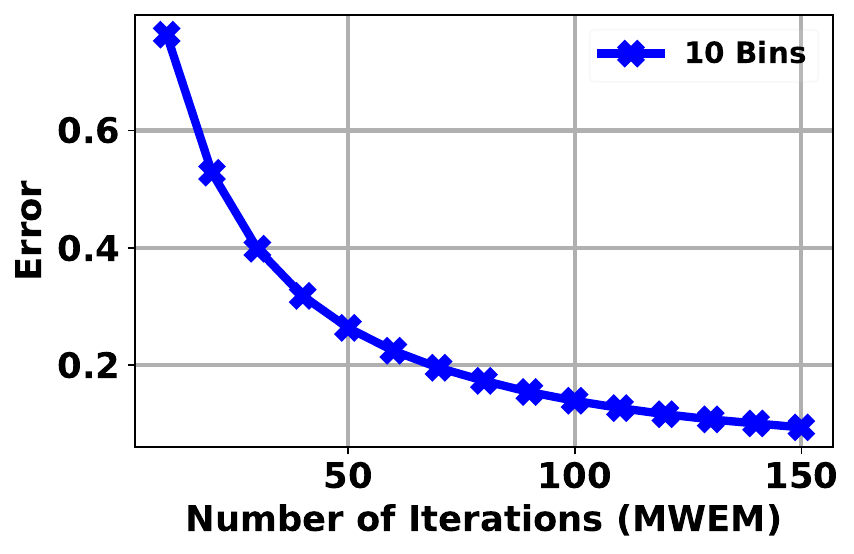}
	\end{subfigure}
	\hfill
	\begin{subfigure}[b]{0.475\linewidth}  
		\centering 
		\includegraphics[width=\textwidth]{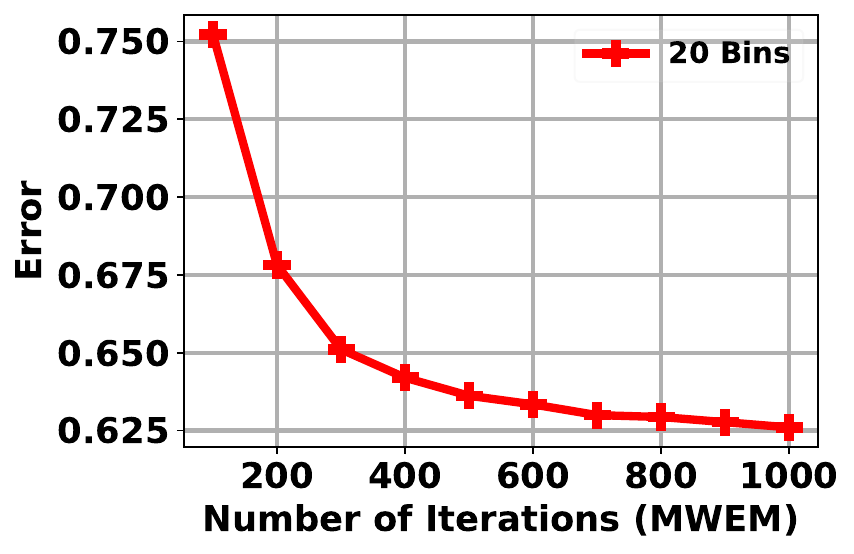}
	\end{subfigure}
	\caption[]
	{\small MWEM error i.e. difference between true and generated distribution, with different number of iterations of the algorithm on a skewed dataset using 10 bins and $\epsilon = 2$.} 
	\label{fig:errors_iter}
\end{figure}

In Figure~\ref{fig:errors_iter}, we have plotted the error in MWEM algorithm, measured as the distance between the generated data distribution and the actual data distribution, against the number of iterations. The code was run without MPC or SGX as it does not affect the measured value (please refer to our discussion on accuracy metrics in~\ref{sec:metrics}). We use a dataset with a skewed distribution for this experiment. This is because since MWEM it initialised with a uniform distribution and intuitively, a skewed dataset may require more iterations for it to converge. We observe that the error continues to fall even at 140 iterations where it is around 0.1, suggesting that there is room for further convergence. In the MPC only setting, we had observed in Figure~\ref{fig:time_iter} that even at a modest 30 iterations, the amount of time approached 50 minutes and the amount of global data exchanged was 360 GB, suggest that running it for further iterations would be impractical. Indeed if we double the number of bins to 20 (as shown on the right side in Figure~\ref{fig:errors_iter}), we notice the convergence is slower. The appropriate number of bins in a practical implementation would depend on the number of attributes and the domain size but it is reasonable to expect that real world datasets, such as Adult and Titanic that we use in the next section, would preclude us from using a MPC only approach.

\section{Results with Shamir Protocol}\label{appendix:third}
\begin{figure}
	\centering
	\begin{subfigure}[b]{0.475\linewidth}
		\centering
		\includegraphics[width=\textwidth]{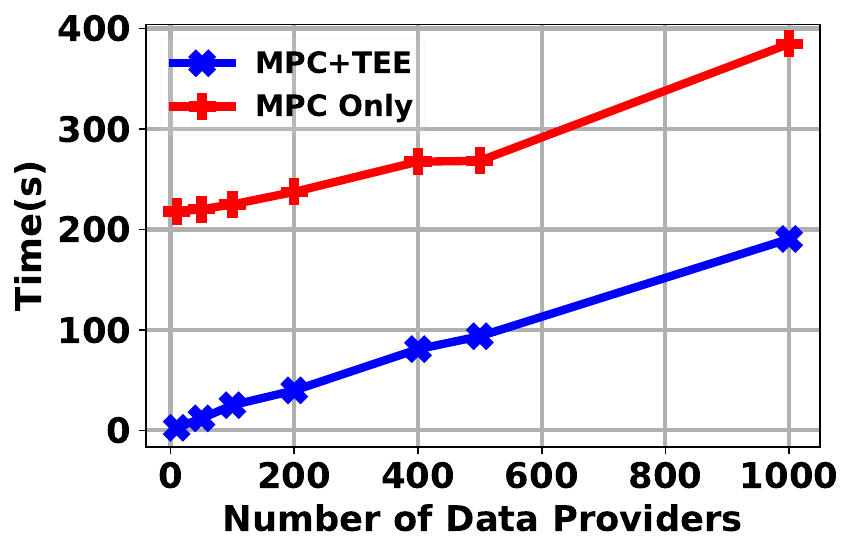}
		\caption[]%
		{{\small Time}}    
	\end{subfigure}
	\hfill
	\begin{subfigure}[b]{0.475\linewidth}  
		\centering 
		\includegraphics[width=\textwidth]{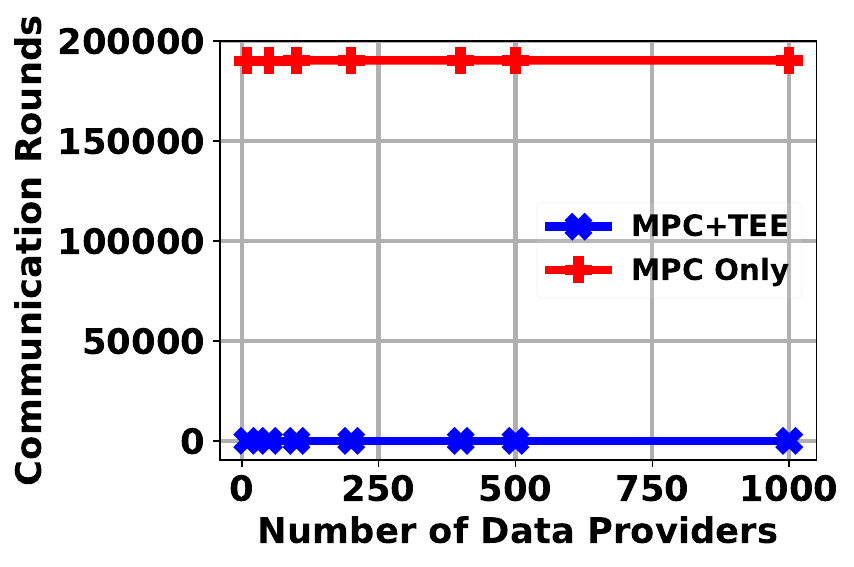}
		\caption[]%
		{{\small Communication Rounds}}    
	\end{subfigure}
	\vskip\baselineskip
	\begin{subfigure}[b]{0.475\linewidth}   
		\centering 
		\includegraphics[width=\textwidth]{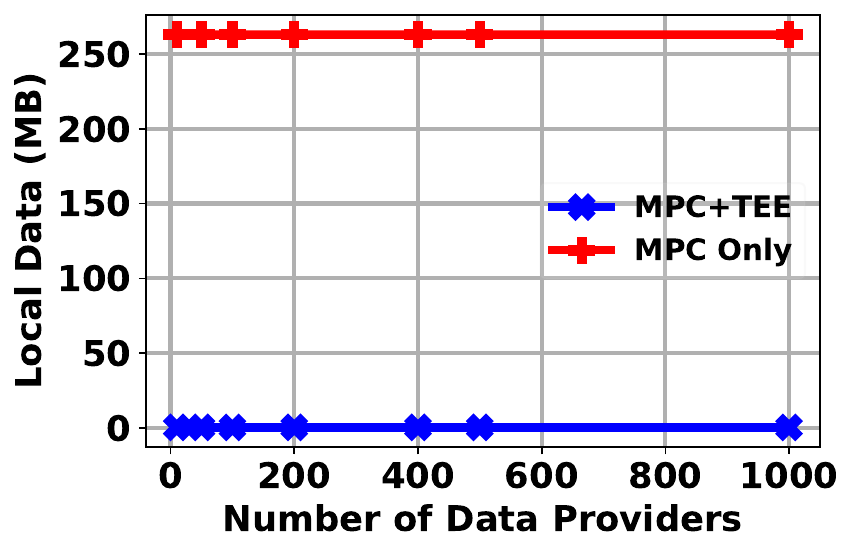}
		\caption[]%
		{{\small Local Data}}    
	\end{subfigure}
	\hfill
	\begin{subfigure}[b]{0.475\linewidth}   
		\centering 
		\includegraphics[width=\textwidth]{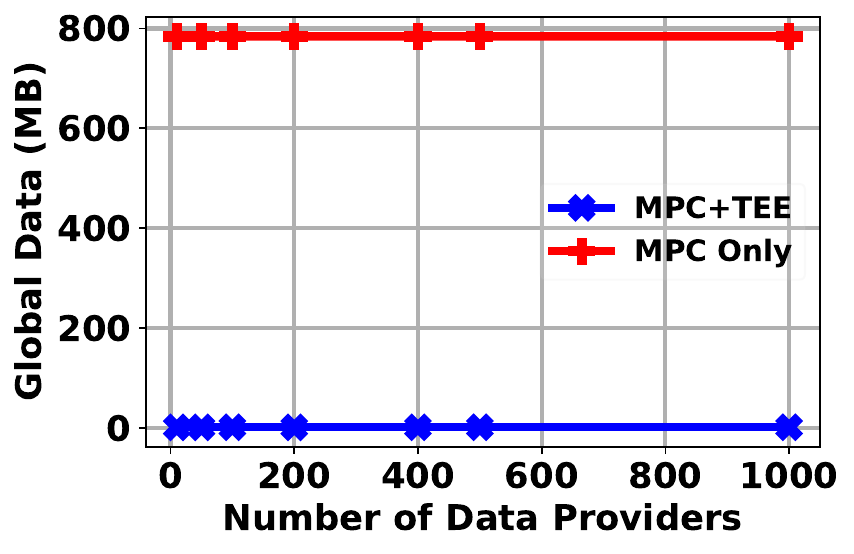}
		\caption[]%
		{{\small Global Data}}    
	\end{subfigure}
	\caption[]
	{\small Comparison of MPC Only and MPC+SGX Approaches [MWEM; SHAMIR protocol, fixed total data (simulated) i.e. 10000 data points divided equally among data providers; 10 bins, $\epsilon = 2$, $T=10 0$.]} 
	\label{fig:fixed_artificial_shamir}
\end{figure}

\begin{figure}
	\centering
	\begin{subfigure}[b]{0.475\linewidth}
		\centering
		\includegraphics[width=\textwidth]{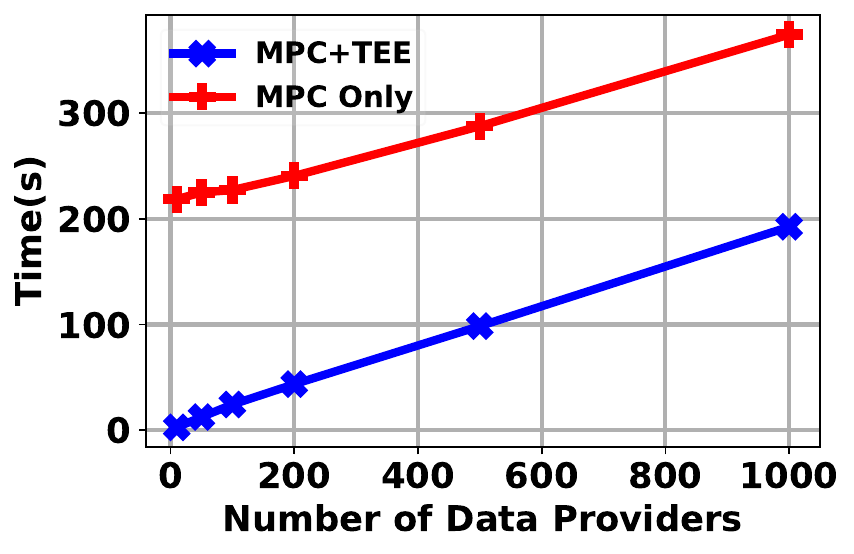}
		\caption[]%
		{{\small Time}}    
	\end{subfigure}
	\hfill
	\begin{subfigure}[b]{0.475\linewidth}  
		\centering 
		\includegraphics[width=\textwidth]{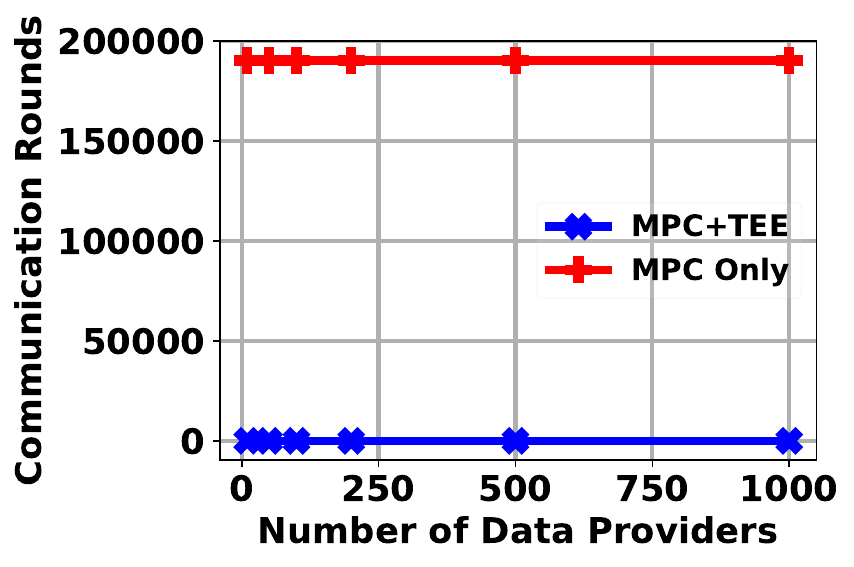}
		\caption[]%
		{{\small Communication Rounds}}   
	\end{subfigure}
	\vskip\baselineskip
	\begin{subfigure}[b]{0.475\linewidth}   
		\centering 
		\includegraphics[width=\textwidth]{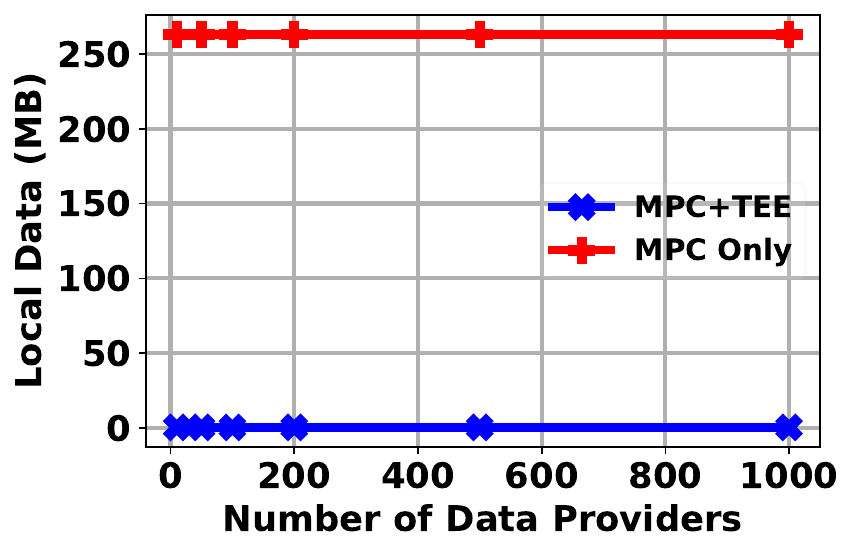}
		\caption[]%
		{{\small Local Data}}    
	\end{subfigure}
	\hfill
	\begin{subfigure}[b]{0.475\linewidth}   
		\centering 
		\includegraphics[width=\textwidth]{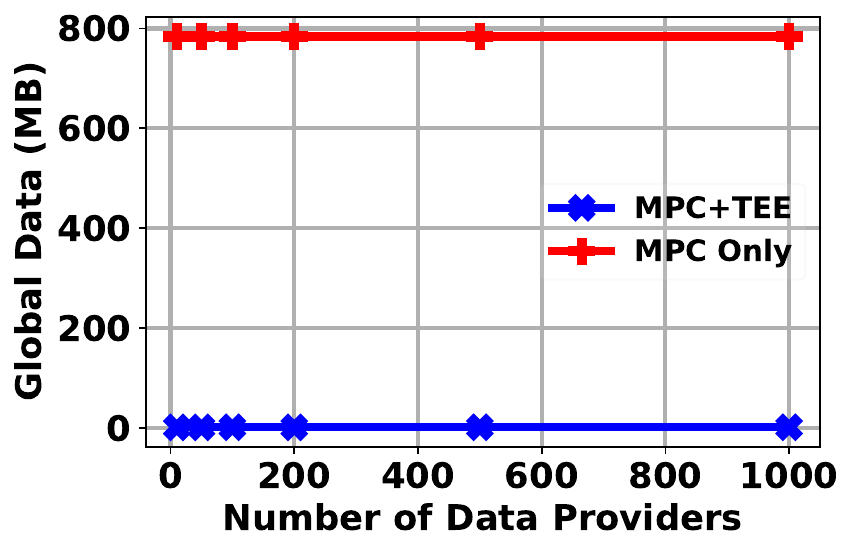}
		\caption[]%
		{{\small Global Data}}    
	\end{subfigure}
	\caption[]
	{\small Comparison of MPC Only and MPC+SGX Approaches [MWEM; SHAMIR protocol, variable total data (simulated) i.e. 100 data points per provider; 10 bins, $\epsilon = 2$, $T=100$.]} 
	\label{fig:fixed_variable_shamir}
\end{figure}

We previously discussed results obtained using the MASCOT MPC protocol (simulating dishonest majority setting). We now report the results for the Shamir MPC protocol to simulate an honest majority setting. Once again, we used the MP-SPDZ~\cite{keller2020mp} implementation for the Shamir MPC protocol. Results are shown in Figures~\ref{fig:fixed_artificial_shamir} and \ref{fig:fixed_variable_shamir} for fixed total data and variable total data settings using the simulated dataset. Being a more efficient protocol due to relaxed adversarial assumptions, we see some clear differences in the behaviour of the metrics under both approaches. Consequently we note that values for time, global and local data are significantly lower than in Mascot. This is despite the fact that the number of communication rounds (around 190000) is not that far off from the rounds in the Mascot setting (approximately 320000), indicating that communication is noticeably cheaper in an honest majority setting. We also observe that the time for both MPC only and MPC+SGX grows linearly at roughly the same rate in this case.

\begin{figure}
	\centering
	\begin{subfigure}[b]{0.475\linewidth}
		\centering
		\includegraphics[width=\textwidth]{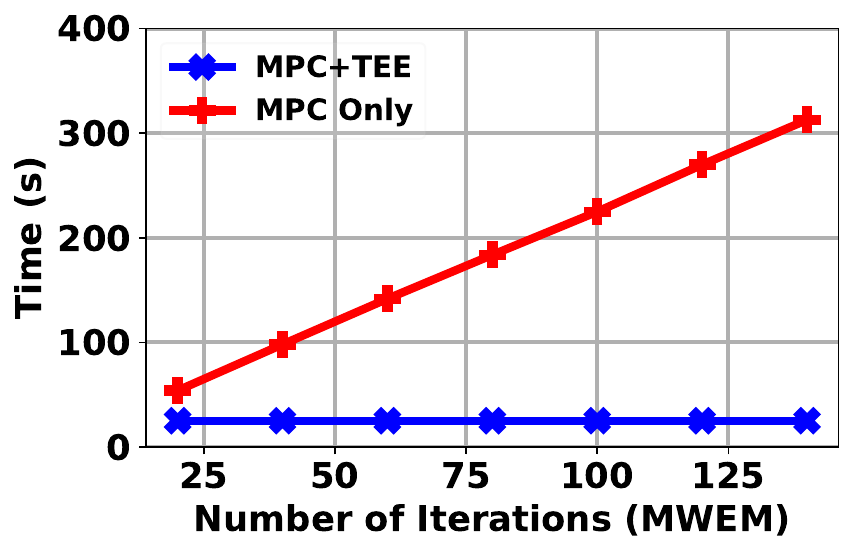}
		\caption[]%
		{{\small Fixed total data (simulated) i.e. 10000 data points divided among 100 data providers.}}    
	\end{subfigure}
	\hfill
	\begin{subfigure}[b]{0.475\linewidth}  
		\centering 
		\includegraphics[width=\textwidth]{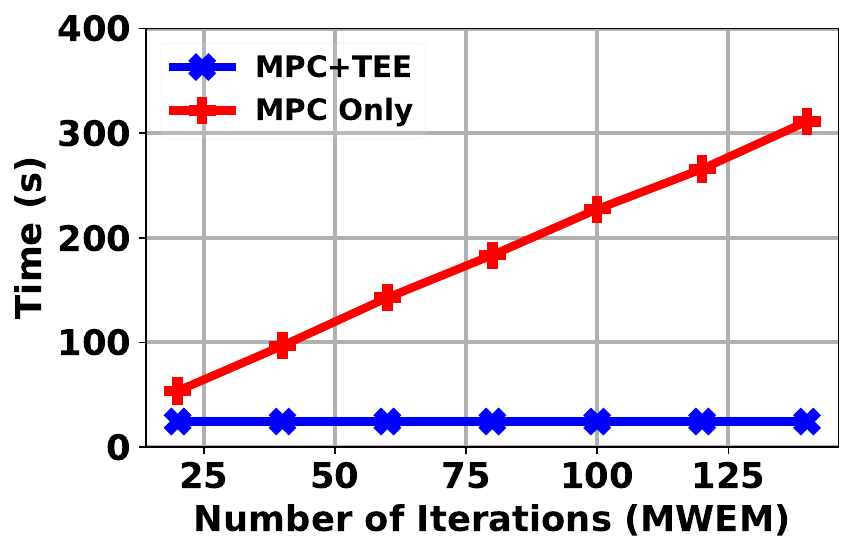}
		\caption[]%
		{{\small Variable Total Data (simulated) i.e. 100 data points per provider.}}   
	\end{subfigure}
	\caption[]
	{\small Comparison of time in MPC Only and MPC+SGX Approaches, for different number of iterations of MWEM. SHAMIR protocol, 100 data providers, 10 bins, $\epsilon = 2$, $T=100$.} 
	\label{fig:time_iter_shamir}
\end{figure}

We observe in figure \ref{fig:time_iter_shamir} that time taken grows much slower in the hybrid MPC+SGX approach, with the number of iterations of the MWEM algorithm. A similar trend (not included as a separate plot in the paper to avoid repetition) was observed for local and global data as well, with global data exceeding 1 GB at 140 iterations in the MPC only approach.

\section{Benchmarking with PGM on Titanic dataset}\label{appendix:forth}
\begin{figure}[H]
	\centering
	\begin{subfigure}[b]{0.475\linewidth}
		\centering
		\includegraphics[width=\textwidth]{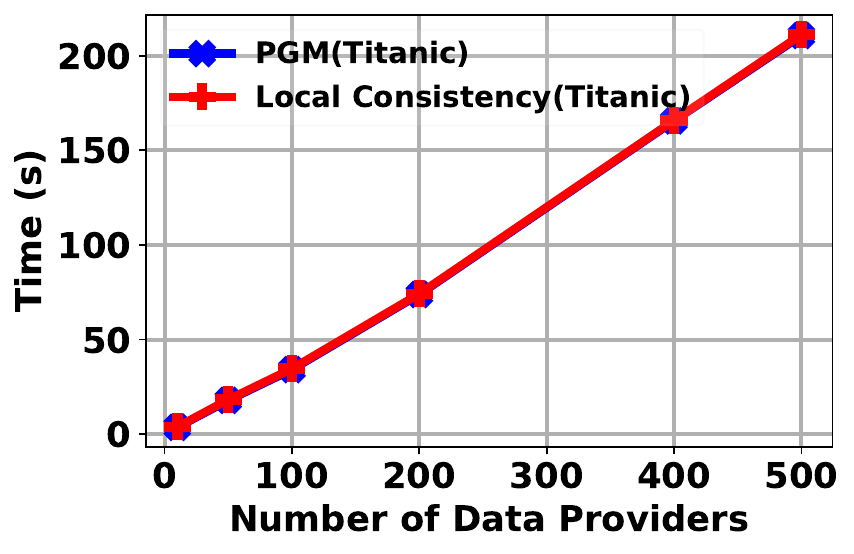}
		\caption[]%
		{{\small Time}}    
		\label{titanic_time}
	\end{subfigure}
	\hfill
	\begin{subfigure}[b]{0.475\linewidth}  
		\centering 
		\includegraphics[width=\textwidth]{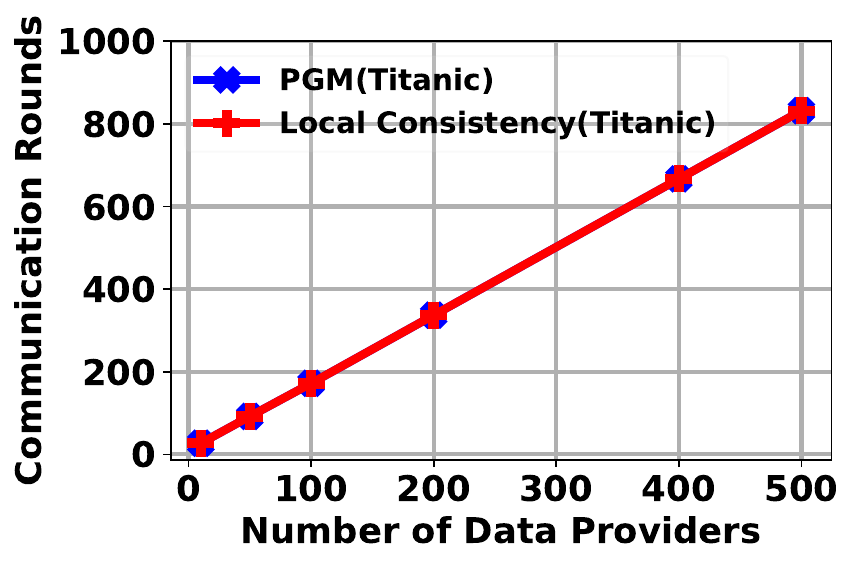}
		\caption[]%
		{{\small Communication Rounds}}    
	\end{subfigure}
	\vskip\baselineskip
	\begin{subfigure}[b]{0.475\linewidth}   
		\centering 
		\includegraphics[width=\textwidth]{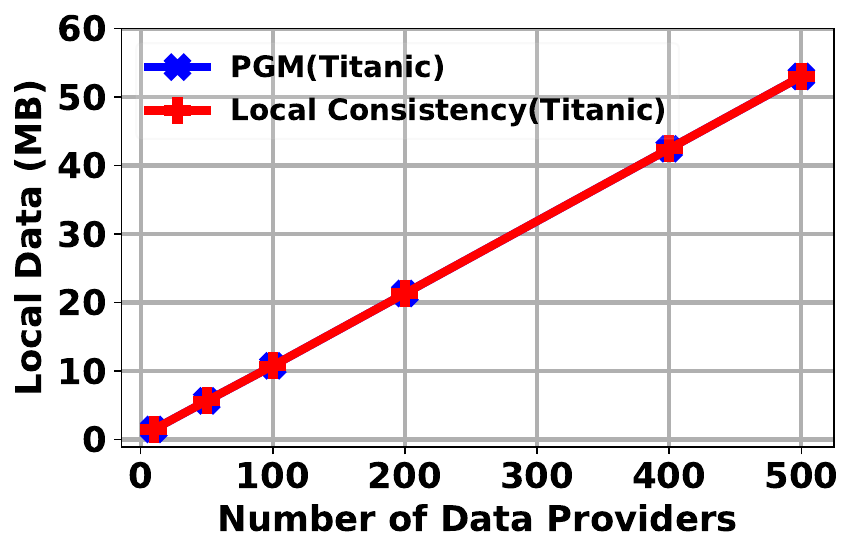}
		\caption[]%
		{{\small Local Data}}    
	\end{subfigure}
	\hfill
	\begin{subfigure}[b]{0.475\linewidth}   
		\centering 
		\includegraphics[width=\textwidth]{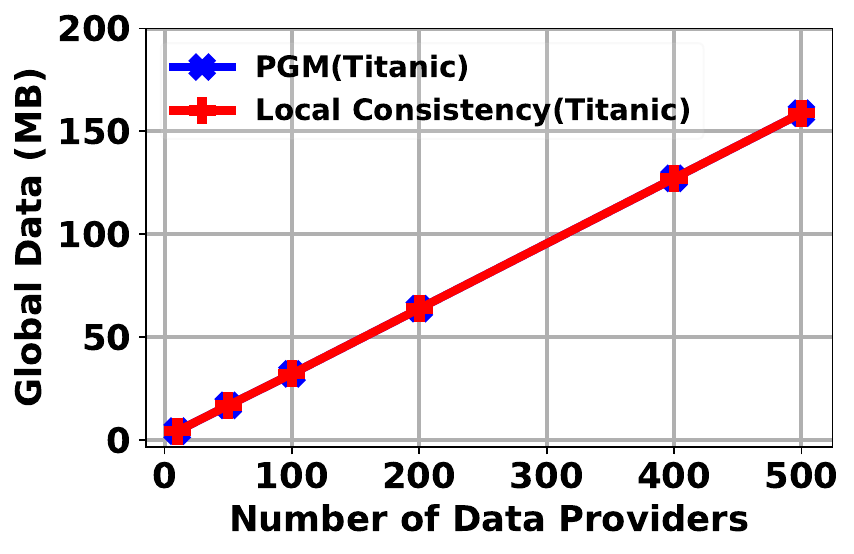}
		\caption[]%
		{{\small Global Data}}    
	\end{subfigure}
	\caption[]
	{\small Performance of the MPC+SGX approach using PGM and Local Consistency generation algorithms (30 iterations); Titanic dataset; SHAMIR protocol, one data point per provider.]} 
	\label{fig:titanic}
\end{figure}

We also benchmark the performance of the MPC+SGX approach  with PGM on the Titanic dataset using the Shamir protocol. We consider all two-dimensional marginals over the 7 attributes resulting in 21 histograms in total. Again PGM is run for 30 iterations. Each data provider represents a single record from Titanic. Results are shown in Figure~\ref{fig:titanic}. Despite considering a greater number of marginals than Adult, we see that it scales better on Titanic across all metrics. This can be attributed to the larger domain size of attributes. As seen in figure \ref{titanic_time} the time taken for 500 data providers is around 210 seconds, compared to over 660 seconds in \ref{adult_time} for Adult.

Figures~\ref{fig:adult} and \ref{fig:titanic} also show the performance using Local Consistency method. While the differences are not significant to be visible in the figures, we observed that on both the Adult and the distributed Adult dataset, Local Consistency based generation takes approximately 30 seconds less time than PGM. On the other hand in the Titanic dataset, Local Consistency based inference takes about 1-1.5 seconds longer than PGM. We also observed, for example, that Local Consistency does not run into the same memory limitations that we encountered when considering more marginals in PGM. The communication rounds, local data and global data does not vary between PGM and Local Consistency in our experiments due to the fact that steps executed in MPC do not differ in the two algorithms.

\end{document}